\documentclass[showpacs,prl,amsmath,amssymb,superscriptaddress,nobibnotes,twocolumn, preprintnumbers]{revtex4-1}
\usepackage{graphicx}
\usepackage{endnotes}
\usepackage{bm}
\usepackage{epsfig}
\usepackage{amsmath}

\newcommand{\ie}{{\it i.e.}}

\newcommand{\SrRh}{Sr$_{3}$Rh$_4$Sn$_{13}$}
\newcommand{\CaSrRh}{(Ca$_{0.5}$Sr$_{0.5}$)$_3$Rh$_4$Sn$_{13}$}
\newcommand{\CaRh}{Ca$_{3}$Rh$_4$Sn$_{13}$}

\begin{document}


\title{Ambient Pressure Structural Quantum Critical Point in the Phase Diagram of (Ca$_x$Sr$_{1-x}$)$_3$Rh$_4$Sn$_{13}$ }

\author{S. K. Goh}
\email{skgoh@phy.cuhk.edu.hk}
\affiliation{Department of Physics, The Chinese University of Hong Kong, Shatin, New Territories, Hong Kong, China}
\affiliation{Cavendish Laboratory, University of Cambridge, J. J. Thomson Avenue, Cambridge CB3 0HE, United Kingdom}

\author{D. A. Tompsett}
\affiliation{Department of Chemistry, Imperial College London, South Kensington SW7 2AZ, United Kingdom}

\author{P. J. Saines}
\affiliation{Department of Chemistry, University of Oxford, Inorganic Chemistry Laboratory, South Parks Road, Oxford OX1 3QR, United Kingdom}

\author{H. C. Chang}
\affiliation{Cavendish Laboratory, University of Cambridge, J. J. Thomson Avenue, Cambridge CB3 0HE, United Kingdom}

\author{T. Matsumoto}
\author{M. Imai}
\author{K. Yoshimura}
\affiliation{Department of Chemistry, Graduate School of Science, Kyoto University, Kyoto 606-8502, Japan}

\author{F. M. Grosche}
\affiliation{Cavendish Laboratory, University of Cambridge, J. J. Thomson Avenue, Cambridge CB3 0HE, United Kingdom}
\date{\today}


\begin{abstract}
The quasi-skutterudite superconductor Sr$_3$Rh$_4$Sn$_{13}$ features a pronounced anomaly in electrical resistivity at $T^*\sim$138~K. We show that the anomaly is caused by a second-order structural transition, which can be tuned to 0~K by applying physical pressure and chemical pressure via the substitution of Ca for Sr. A broad superconducting dome is centred around the structural quantum critical point. Detailed analysis of the tuning parameter dependence of $T^*$ as well as insights from lattice dynamics calculations strongly support the existence of a structural quantum critical point at ambient pressure when the fraction of Ca is 0.9 (\ie, $x_c=0.9$). This establishes (Ca$_x$Sr$_{1-x}$)$_3$Rh$_4$Sn$_{13}$ series as an important system for exploring the physics of structural quantum criticality without the need of applying high pressures.

\end{abstract}


\maketitle


Quantum criticality associated with a second-order phase transition at zero temperature has been a fruitful concept in the search for superconductivity. Particularly clear examples include the heavy fermion \cite{Mathur98, Gegenwart08} and iron-based systems \cite{Paglione10, Ishida09, Hashimoto12, Shibauchi14}, where superconductivity is stabilized in the vicinity of an antiferromagnetic quantum critical point (QCP). In addition to magnetic instabilities, structural instabilities also have an intricate influence on superconductivity. However, clean systems with continuous structural transition tunable by external parameters are relatively rare, and for the few cases studied in recent years, the structural order is found in the vicinity of magnetic order \cite{Yoshizawa12}, making the notion of structural quantum criticality and its influence on the superconductivity less advanced than the magnetic counterpart. 

Superconducting stannides \cite{Remeika80, Espinosa80}, including A$_3$T$_4$Sn$_{13}$ systems where A=La,Sr,Ca and T=Ir,Rh, have recently been reexamined using a wide range of probes \cite{Yang10, Kase11, Wang12, Klintberg12, Zhou12, Gerber13, Liu13, Slebarski14, Tompsett14, Kuo14, Biswas14, Sarkar14, Fang14}. \SrRh, a member of this stoichiometry family, is a strong-coupling superconductor with a critical temperature $T_c$~$\approx$4.7~K. The superconducting gap function $\Delta$ is nodeless, with $2\Delta/k_BT_c=3.87$ \cite{Kase11}. In addition to superconductivity, the normal state of \SrRh\ features another instability which sets in at $T^*\approx138$~K. In \CaRh, which is also a superconductor with $T_c$=7.8~K, no signature of the anomaly associated with $T^*$ was observed. The substitution series (Ca$_x$Sr$_{1-x}$)$_3$Rh$_4$Sn$_{13}$ is readily available, allowing us to track the evolution of $T^*$ and its interplay with superconductivity when Ca is partially replaced by Sr, which simulates a negative chemical pressure, as well as when positive physical pressure is applied directly on the crystals.

A recent investigation of the related series (Ca$_x$Sr$_{1-x}$)$_3$Ir$_4$Sn$_{13}$ revealed the existence of a structural quantum phase transition at 18 kbar in Ca$_3$Ir$_4$Sn$_{13}$ \cite{Klintberg12}. Since Ca$_3$Ir$_4$Sn$_{13}$ sits at the limit of the chemical pressure in this series, physical pressure has to be applied in order to access and study the physics of structural quantum phase transition, thus restricting the accessibility of many experimental probes. In this Letter, we argue that the (Ca$_x$Sr$_{1-x}$)$_3$Rh$_4$Sn$_{13}$ series is an excellent alternative for the investigation of structural criticality. Here, we
show that $T^*$ is a second-order structural transition temperature which can be suppressed with a suitable combination of physical and chemical pressures. Importantly, $T^*$ can be driven to 0~K for (Ca$_{0.9}$Sr$_{0.1}$)$_3$Rh$_4$Sn$_{13}$ (\ie,  $x=0.9$) without applying physical pressure. In the vicinity of $x=0.9$, the structural QCP, a broad superconducting dome is observed, giving rise to a phase diagram reminiscent of cases featuring an antiferromagnetic QCP.


Single crystals of (Ca$_x$Sr$_{1-x}$)$_3$Rh$_4$Sn$_{13}$ were synthesized by the Sn flux method using similar parameters as described in ref. \cite{Yang10}. High pressure electrical resistivity, $\rho(T)$, was measured using a four-wire configuration in a piston-cylinder cell with Daphne 7373 as the pressure transmitting medium. The pressure achieved was determined using the superconducting transition of lead. Heat capacity, $C_p(T)$, was measured using a standard pulse relaxation method. Both $C_p(T)$ and $\rho(T)$ were measured in a Physical Property Measurement System (Quantum Design). Single crystal XRD measurements were performed, using an Oxford Diffraction Supernova A utilising MoK$\alpha$ radiation and equipped with a Oxford Cryostream Plus. Data integration and analytical absorption corrections were performed with Crysalis Pro \cite{CrysAlisPro}. Structures were solved using direct methods and refined against $|F|^2$ using Shelx-97 \cite{Sheldrick} via the Olex-2 \cite{Dolomanov09} interface. The electronic structure has been calculated using the Local Density Approximation (LDA). The VASP \cite{Kresse96} code was employed using projector-augmented wave potentials. A planewave cutoff of 300 eV was used and a minimum 4x4x4 $k$-point grid was set up, which was scaled appropriately for supercells. The phonon spectrum was calculated by using PHONOPY \cite{phonopy1} in conjunction with VASP. All calculations were non-spin polarized.
\begin{figure}[!t]\centering
      \resizebox{8.5cm}{!}{
              \includegraphics{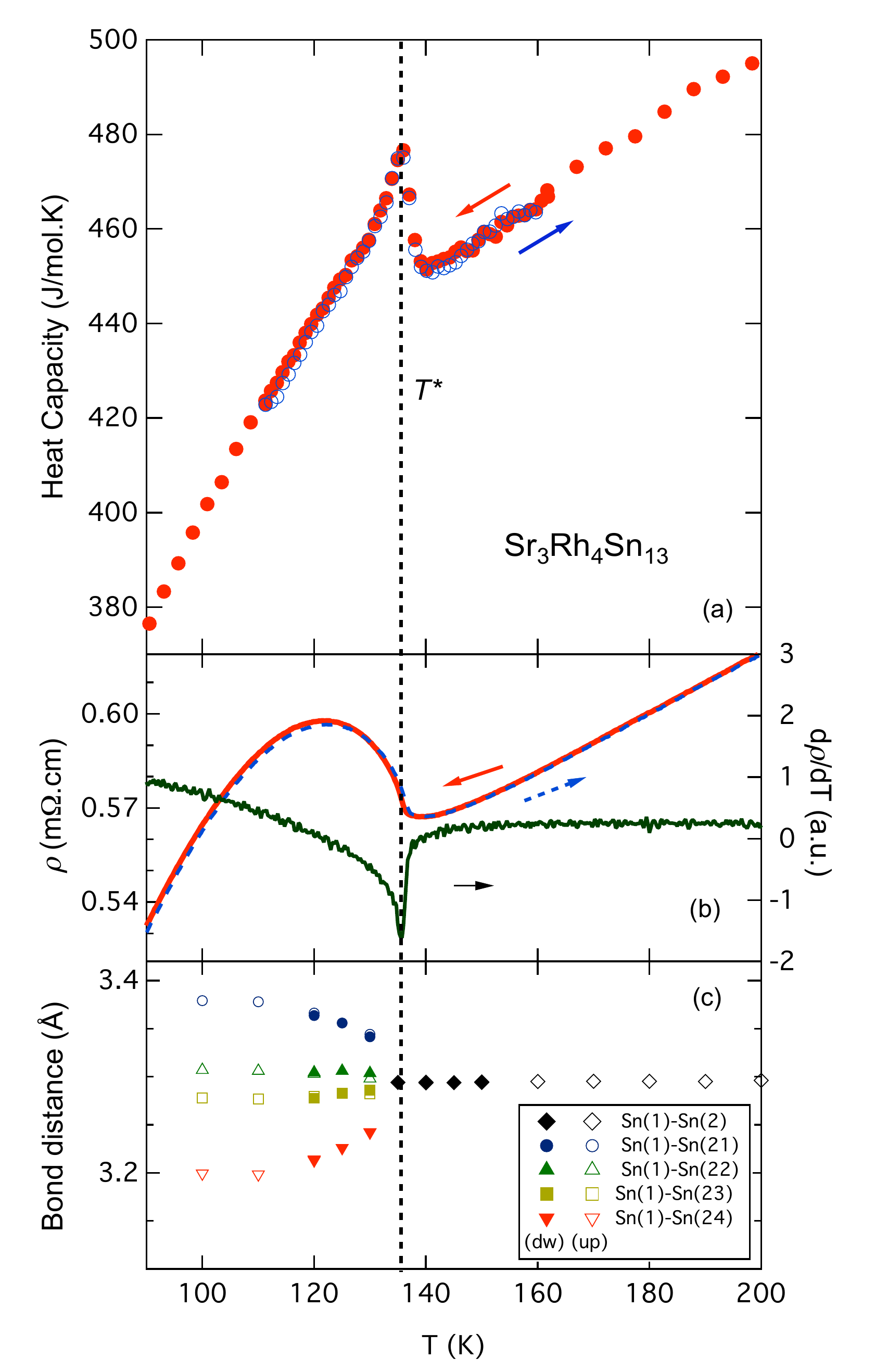}}                				
              \caption{\label{fig1} (Color online) The temperature dependence of the (a) heat capacity, (b) electrical resistivity and (c) Sn(1) -- Sn(2) bond distances for \SrRh\ near $T^*$. In (a) and (c), the closed (open) symbols denote the data collected on cooling (warming). In (b), the solid (broken) line is the cooling (warming) curve. All these point to the absence of hysteresis at $T^*$. The temperature derivative of the electrical resistivity in (b) provides a means to define $T^*$.}
\end{figure}

The normal state of \SrRh\ features a phase transition at $T^*\approx$138~K, which is detectable in a wide range of physical quantities including the heat capacity (Fig. \ref{fig1}(a)) and electrical resistivity (Fig. \ref{fig1}(b)). The minimum in $d\rho/dT$ is used to define $T^*$, as indicated in Fig. \ref{fig1}(b). The measurements of both $C_p(T)$ and $\rho(T)$ have been carried out on cooling down and warming up, at the same sweep rates. The cooling and warming curves overlap perfectly, showing the absence of hysteresis. In addition, $C_p(T)$ near $T^*$ exhibits a lambda-like jump. These observations indicate that $T^*$ corresponds to a transition temperature of a second-order or only very weakly first-order phase transition.

Single crystal X-ray diffraction measurements at 300 K confirmed that \SrRh\ adopts a $Pm\bar{3}n$ cubic structure, the $I$ phase, with $a = 9.80610(18)$~\AA. Below $T^*$, additional weak reflections are observed that are consistent with a phase transition to a body-centred cubic structure, the $I'$ phase, with a lattice parameter twice that of the high temperature phase (Section C, Ref. \cite{SUPP}). An acceptable refinement could only be achieved in space group $I\bar{4}3d$ with $a = 19.5714(3)$~\AA\ at 100 K ($R$1$_{obs}$, $wR$2$_{obs}$ and $\chi^2$ of 3.1~\%, 8.9~\% and 1.1~\%). The structural distortion is similar to that observed in Sr$_3$Ir$_4$Sn$_{13}$ at 147 K \cite{Klintberg12, Tompsett14}, with the bond distances in the Sn(1)Sn(2)$_{12}$ icosahedra distorting into four groups of three identical bond distances (see Fig. 1c). This occurs in concert with tilting of three-quarters of the RhSn(2)$_6$ trigonal prisms with only those trigonal prisms whose axes are along the (111) direction remaining untilted. The transition temperature remains the same on heating and cooling, consistent with the expected second-order transition between $Pm\bar{3}n$ and $I\bar{4}3d$ space groups, which have a group-subgroup relationship.

Mode decomposition of the low temperature phase was carried out using the ISODISTORT software suite \cite{Campbell06} by comparison with the parent high symmetry phase. It revealed that the active ${\bf q}$-vector is (0.5, 0.5, 0) and its symmetry equivalents. In particular, the structural distortion appears to be primarily driven by the modes belonging to the M$_2-$irreducible representation, primarily causing displacement of the Sn(2) sites along with small movements of the strontium cations. It has previously been shown that the transition between the primitive and body-centred structures is linked to charge transfer between Sn(2) and Sn(1) atoms, which increases the ionicity of the bond between these atoms \cite{Bordet91, Miraglia86}. 

\begin{figure}[!t]\centering
       \resizebox{9cm}{!}{
              \includegraphics{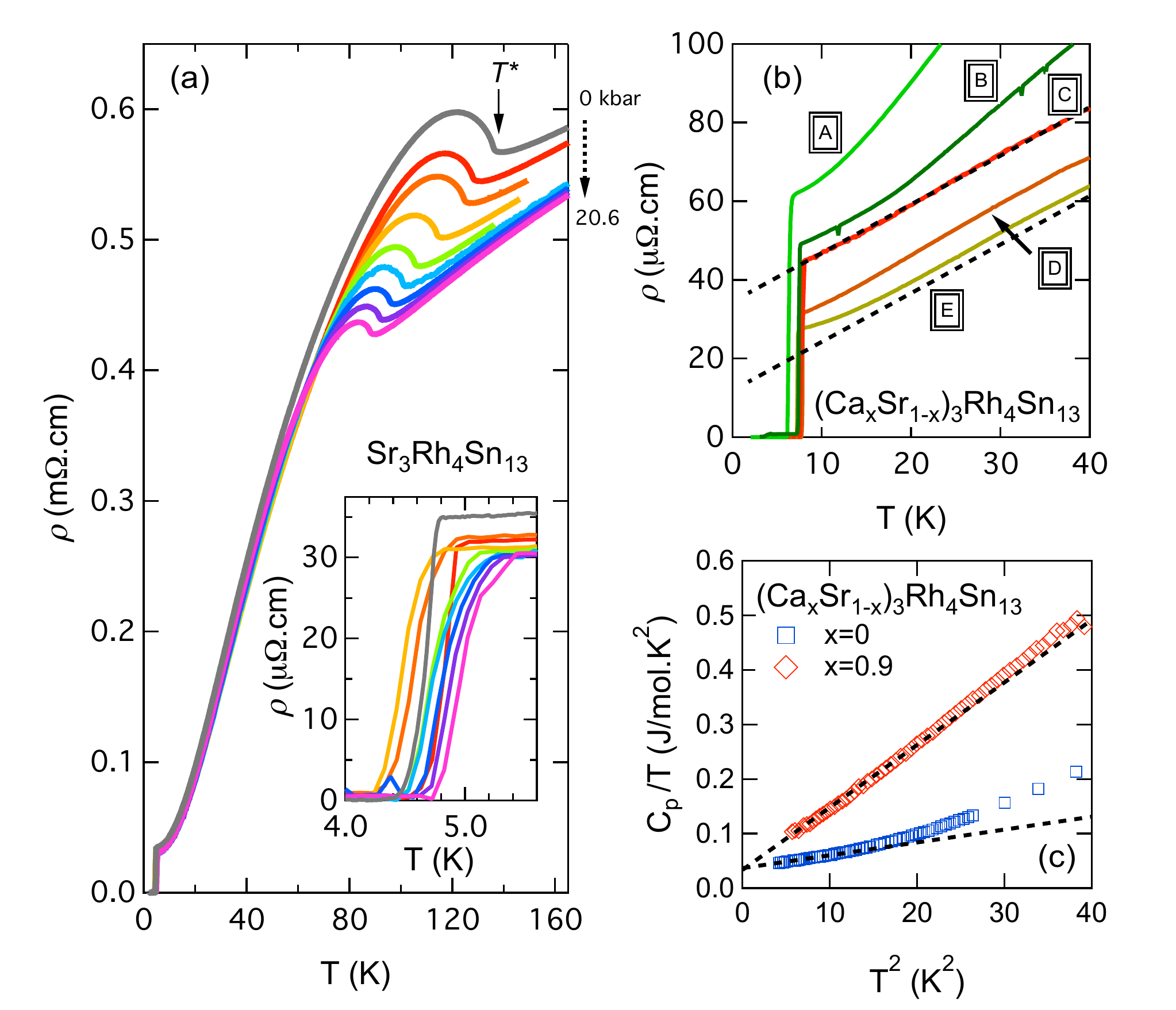}}                				
              \caption{\label{fig2} (Color online) (a) High pressure $\rho(T)$ in \SrRh\ for the determination of $T^*$ and $T_c$ (inset). The pressure values increase sequentially as indicated by the arrows, and are given by 0, 4.0, 6.0, 12.7, 14.0, 15.0, 17.3, 19.1, 20.6 kbar. (b) Representative $\rho(T)$ for the system situated at 5 different locations of the universal phase diagram (see arrows in Fig.~\ref{figpd}), spanning a range which straddles the QCP. The dashed straight lines are guides for the eyes \cite{SCALE}. (c) The normal state $C_p/T$ is plotted against $T^2$ for $x=x_c=0.9$ ({\large $\diamond$}), right at the quantum critical point, and $x=0$ ($\Box$).}
\end{figure}
Having established the origin of the anomaly at $T^*$, we now examine its pressure dependence.  We show in Fig. \ref{fig2}(a) the high pressure $\rho(T)$ of \SrRh\ ($x=0$).  $T^*$ decreases rapidly when hydrostatic pressure is applied. Concurrently, the superconducting transition temperature $T_c$, defined using the resistive mid-point of the transition, changes only slightly under pressure. In \CaRh\ ($x=1$), no signature of $T^*$ was detected for all pressures studied, whilst $T_c$ is higher than that in \SrRh\ and shows a gentle variation with pressures (Figs.~S5--S6, Ref.~\cite{SUPP}).
\begin{figure}[!t]\centering
       \resizebox{7.5cm}{!}{
              \includegraphics{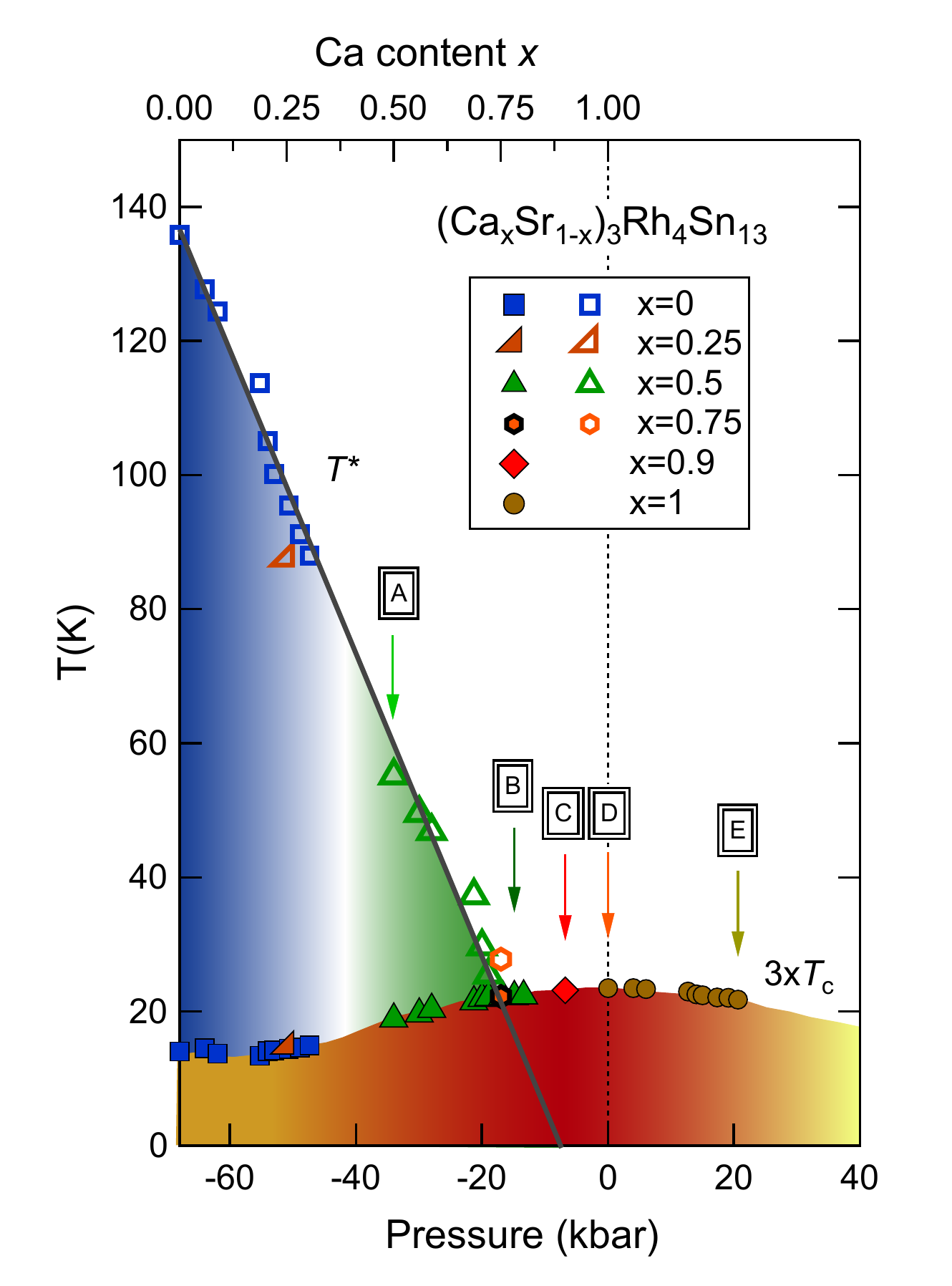}}              				
              \caption{\label{figpd} (Color online) Universal phase diagram showing the pressure dependence of $T^*$ and $T_c$ for (Ca$_x$Sr$_{1-x}$)$_3$Rh$_4$Sn$_{13}$. The origin of the pressure scale is chosen to coincide with $x=1$, as indicated by the dashed vertical line. In this representation, the part of the phase diagram to the left of the dashed line is accessible by varying the calcium content, whereas the part to the right is accessible only by physical pressures. The arrows indicate the positions in the parameter space where $\rho(T)$ curves are presented in Fig. \ref{fig2}(b)}
\end{figure}

Both \SrRh\ and \CaRh\ adopt the $Pm\bar{3}n$ structure at room temperature; the lattice constant of (Ca$_x$Sr$_{1-x}$)$_3$Rh$_4$Sn$_{13}$ shrinks linearly and uniformly with increasing calcium content $x$, obeying Vegard's law (Fig.~S1, Ref.~\cite{SUPP}). Therefore, the substitution of Ca for Sr simulates the effect of hydrostatic pressure, and \CaRh\ can be regarded as the high pressure analogue of \SrRh. Detailed analysis of all $T^*$ and $T_c$ data suggests that $\Delta x=1$ corresponds to $\Delta p=68$~kbar (see Section A, Ref.~\cite{SUPP}). Placing \SrRh\ at $-68$~kbar and \CaRh\ at the origin of the pressure scale, we construct the universal phase diagram (Fig.~\ref{figpd}) which illustrates the interplay between the structural order and superconductivity driven by both the physical pressure and the chemical pressure.
The universal phase diagram is reminiscent of the cases where  superconductivity is stabilized on the border of magnetism (e.g. \cite{Mathur98, Gegenwart08,Paglione10, Ishida09, Hashimoto12, Shibauchi14}). The crucial difference here is that it is a structural transition which is suppressed by the nonthermal tuning parameter. $T^*$ extrapolates to 0~K at $p_c=-6.8$~kbar relative to \CaRh, or equivalently, at $x_c=0.9$. If the structural transition remains second-order, or weakly first-order, a structural QCP at $x_c$ or $p_c$ is expected. Thus, the phase diagram of (Ca$_x$Sr$_{1-x}$)$_3$Rh$_4$Sn$_{13}$ features a QCP which is accessible at ambient pressure by fine tuning the Ca content.

When a crystalline system undergoes a second-order displacive structural transition, the relevant phonon mode must go soft at the transition temperature $T^*$. The existence of a phonon mode at $Q\neq0$ such that $\omega_Q\rightarrow0$ implies the emergence of a new periodicity. When $T^*$ is tuned to 0~K, the associated softening of the phonon mode occurs at 0~K. This gives rise to additional low-lying phonon modes, which can be excited at low temperatures. Using the generalised Bloch-Gr\"{u}neisen expression \cite{Gauzzi07, Lortz06}, the contribution to the electrical resistivity due to phonon scattering can be written as:
\begin{equation}
\Delta \rho_{ph}(T)\propto\sum_{\bf q}\alpha^2_{(tr){\bf q}}\frac{x_{\bf q}e^{x_{\bf q}}}{(e^{x_{\bf q}}-1)^2} 
\approx \sum_{\bf q}\alpha^2_{(tr){\bf q}}\left(\frac{k_B}{\hbar\omega_{\bf q}}\right) T,
\nonumber
\end{equation}
where $\alpha^2_{(tr){\bf q}}$ is a ${\bf q}$-dependent Fermi surface average of the electron-phonon interaction, $\omega_{\bf q}$ is the phonon frequency and $x_{\bf q}=\hbar\omega_{\bf q}/k_BT$. The approximation in the last part of the expression is applicable when $\hbar\omega_{\bf q}\ll k_BT$, \ie, the softening of the phonon frequency relative to the thermal energy. Therefore, the phononic contribution to the electrical resistivity is linear in $T$ when $\hbar\omega_{\bf q}\ll k_BT$, the degree of contribution is enhanced with softer $\omega_{\bf q}$. 

Fig. \ref{fig2}(b) shows five representative traces of $\rho(T)$ whose composition and pressure values are chosen to straddle $x_c=0.9$ ($p_c=-6.8$~kbar), the location where $T^*\rightarrow0$~K, and their positions in the universal phase diagram are indicated by the arrows in Fig. \ref{figpd}. Far below $x_c$, $\rho(T)$ exhibits a large curvature just above $T_c$. However, at $x_c$, a distinct $T$-linear resistivity is observed over a wide temperature range (Curve C). According to our treatment in the preceding paragraph, this observation at low temperatures suggests the existence of a very soft phonon mode, so as to satisfy $\hbar\omega_{\bf q}\ll k_BT$. This is consistent with the scenario of a second-order structural quantum phase transition at $x_c$.
Moving away from $x_c$, the curvature gradually builds up again, as evidenced in the  $\rho(T)$ traces labelled D and E (see also Fig.~S7, Ref.~\cite{SUPP} for intermediate $\rho(T)$ between D and E).

\begin{figure}[!t]\centering
       \resizebox{8.5cm}{!}{
              \includegraphics{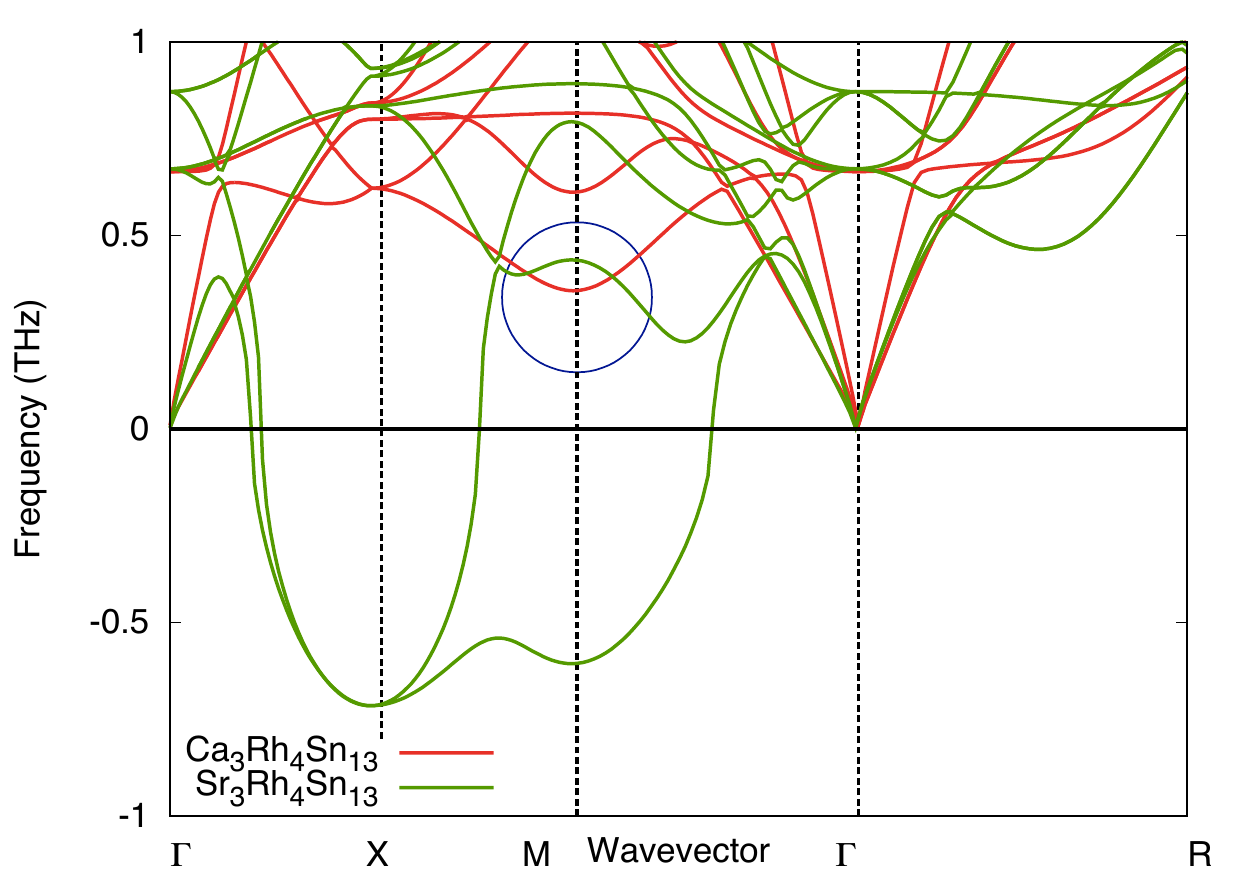}  }           				
              \caption{\label{dispersion} (Color online) Dispersion relation $\omega(q)$ for \SrRh\ and \CaRh\ at $T=0$. The circle highlights the presence of an anomalously soft phonon mode in \CaRh\ near M. The negative part of the frequency axis denotes imaginary phonon mode frequencies.}
\end{figure}

We can gain further insight by calculating the phonon dispersion relation for \SrRh, which is below $x_c$, and \CaRh, which according to our experimental results is above $x_c$. In Fig. \ref{dispersion} the low energy modes of the phonon spectrum are shown for \CaRh\ and \SrRh. The depicted points along the high symmetry lines lie at coordinates X(0.5,0,0), M(0.5,0.5,0) and R(0.5,0.5,0.5) in the Brillouin zone corresponding to the simple cubic lattice. While \SrRh\ with its larger lattice parameter possesses imaginary mode frequencies, indicating a lattice instability, for \CaRh\ these modes have hardened to yield real frequencies. This is consistent with the experimentally observed phase diagram, in which \SrRh\ undergoes a structural transition at $\sim$138~K while for \CaRh\ the transition is absent. However, as highlighted by the circle in the figure, \CaRh\ retains a soft mode of approximately 0.35 THz at M, the wavevector associated with the structural transition in \SrRh. This suggests that \CaRh\ is near to the critical end point for the structural phase transition, and the calculations lend further support to our conclusion that a structural QCP is reachable in the (Ca$_x$Sr$_{1-x}$)$_3$Rh$_4$Sn$_{13}$ series by tuning the Ca content.

When a QCP located at $x_c$ is approached, the order parameter correlation length $\xi$ and correlation time $\xi_\tau$ diverge according to $\xi\sim|x-x_c|^{-v}$ and $\xi_\tau\sim\xi^z$ \cite{Shibauchi14, Sachdevbook}. The dynamical exponent $z$ can be determined from the dispersion relation at $x_c$: $\omega\propto q'^z$. We can analyze our system by approaching the critical concentration $x_c$ from the quantum disordered side. The dispersion relation for \CaRh\ ($x>x_c$) around the minimum at M (see Fig. \ref{dispersion}) can be Taylor-expanded as $\omega\propto\sqrt{c^2q'^2+\Delta^2}$, where $q'$ is the wavevector measured from M and $\Delta$ is the gap between the soft mode frequency at M and $\omega=0$. With the alloying of Sr, $\Delta$ decreases and it eventually vanishes at $x_c$, triggering a structural quantum phase transition. At $x_c$, the dispersion relation around M becomes $\omega\propto q'$, analogous to an acoustic branch with slope $c$. Consequently, $z$ is unity and the effective dimension for quantum criticality in our system $d_{\text{eff}}=d+z=4$, where $d$ is the spatial dimension \cite{Shibauchi14,Sachdevbook}. 

The additional phonon branch with a linear dispersion relation at the QCP strongly affects the low temperature heat capacity. Analogous to Debye's treatment, the contribution of this additional branch to the heat capacity is proportional to $T^3$. In Fig. \ref{fig2}(c), we plot the normal state $C_p/T$ versus $T^2$ for $x=0$ and $x=x_c=0.9$. The low temperature part of the data can be described by $C_p/T=\gamma+\beta T^2$ (dashed lines), where $\gamma$ is the coefficient governing the electronic contribution to the heat capacity and the slope $\beta$ is related to the phonon contribution due to linear $\omega(q)$. While $\gamma$ is more or less the same in both compositions, $\beta$ at the QCP is $\sim$4.6 times larger than that at $x=0$, which is far away from the QCP. This additional $T^3$ contribution at the QCP is consistent with the model we proposed in the preceding paragraph.


In summary, we have established the $T-p$ phase diagram of (Ca$_x$Sr$_{1-x}$)$_3$Rh$_4$Sn$_{13}$. We show that a second-order structural transition temperature $T^*$ can be tuned to 0~K by a suitable combination of $p$ and $x$. When $T^*\rightarrow0$, a linear-in-$T$ electrical resistivity is observed, a significant $T^3$ contribution to the heat capacity is recorded, and superconducting transition temperature peaks near this pressure/composition. Lattice dynamics calculations reveal the existence of imaginary phonon frequencies in \SrRh, indicative of structural instabilities. These imaginary phonon mode frequencies completely disappear in \CaRh. Our combined experimental and computational effort highlights a particularly important feature of this series, in that $T^*\rightarrow0$ can be achieved in $x=0.9$ without the need of applying pressure, which opens up a new avenue for detailed investigation of the physics of structural quantum criticality.

\begin{acknowledgments}{\bf Acknowledgement.} We acknowledge funding support from the EPSRC (EP/L000202), Trinity College (Cambridge), Grants-in-Aid from MEXT (22350029 and 23550152), CUHK (Startup Grant, Direct Grant No. 4053071) and Glasstone Bequest, Oxford, as well as the provision of ARCHER supercomputer resources by the Materials Chemistry Consortium.
\end{acknowledgments}
  


\begin{thebibliography}{34}
\bibitem{SCALE}
Curves A and B correspond to $\rho(T)$ in $x=0.5$ at ambient pressure and 19.1~kbar, respectively. Curves D and E correspond to $\rho(T)$ in $x=1.0$ at ambient pressure and 20.6~kbar, respectively. Curve C is for $x_c=0.9$ at ambient pressure. For clarity, the resistivities are multiplied by $0.5$ for ``A" and ``B", and by $1.5$ for ``D" and ``E". Curve C is not scaled.
\end{thebibliography}


\end{document}